\documentclass[11pt]{article}
\usepackage{erice,epsfig}

\def\Journal#1#2#3#4{{#1} {\bf #2}, #3 (#4)}

\def\00186{\em Matters Grav.}
\def\APJSA{\em Astrophys. J. Supp.} 
\def\CMP{\em Commun. Math. Phys.}
\def\DANKA{\em Dokl. Akad. Nauk Ser. Fiz.} 
\def\EPHJA{\em Eur. Phys. J.}
\def\FPYKA{\em Fortsch. Phys.}
\def\IMPAE{\em Int. J. Mod. Phys.}
\def\JAP{\em J. Appl. Phys.}
\def\JTPLA{\em JETP Lett.}
\def\NATUA{\em Nature}

\def\NUPHA{\em Nucl. Phys.}
\def\NPPS{\em Nucl. Phys. Proc. Suppl.} 
\def\OPCOB{\em Opt. Commun.}
\def\PHWOE{\em Phys. World}
\def\PLACB{\em Part. Accel.}
\def\PHLTA{\em Phys. Lett.}
\def\PR{\em Phys. Rev.}
\def\PREP{\em Phys. Rept.} 
\def\PRL{\em Phys. Rev. Lett.}
\def\PRD{{\em Phys. Rev.} D}
\def\PRSLA{{\em Proc. Roy. Soc. Lond.} A}
\def\ZFPRA{\em Pisma Zh. Eksp. Teor. Fiz.}
\def\SCIEA{\em Science}
\def\SJNCA{\em Sov. J. Nucl. Phys.} 
\def\SPHDA{\em Sov. Phys. Dokl.}
\def\SPHJA{\em Sov. Phys. JETP}
\def\SOPUA{\em Sov. Phys. Usp.}
\def\UFNAA{\em Usp. Fiz. Nauk.}
\def\YAFIA{\em Yad. Fiz.}
\def\ZEPYA{\em Z. Phys.} 

\def\ZETFA{\em Zh. Eksp. Teor. Fiz.}

\def\be{\begin{equation}}
\def\ee{\end{equation}}
\def\bea{\begin{eqnarray}}
\def\eea{\end{eqnarray}}

\newcommand{\lwig}{\mbox{\,\raisebox{.3ex}
    {$<$}$\!\!\!\!\!$\raisebox{-.9ex}{$\sim$}\,}}
\newcommand{\gwig}{\mbox{\,\raisebox{.3ex}
    {$>$}$\!\!\!\!\!$\raisebox{-.9ex}{$\sim$}}\,}
\newcommand{\lambdabar}{{\hbox{$\lambda_e$\kern-1.9ex\raise+0.45ex\hbox{--}
\kern+0.2ex}}}

\begin{document}
\vspace*{4cm}
\title{
\vspace*{-8ex}
{\normalsize\rightline{\rm DESY 01-213}\rightline{\rm hep-ph/0112254}}
\vspace*{4ex}
FUNDAMENTAL PHYSICS AT AN X-RAY FREE ELECTRON LASER\footnote{Invited talk at the 
{\em Workshop on Electromagnetic Probes of Fundamental Physics}, Erice, Italy, October 2001.}}

\author{ A. RINGWALD }

\address{Deutsches Elektronen-Synchrotron DESY, Notkestra\ss e 85,\\
D-22607 Hamburg, Germany}

\maketitle\abstracts{
X-ray free electron lasers (FELs) have been proposed  to be constructed 
both at SLAC in the form of the so-called Linac Coherent Light 
Source   as well as at DESY, where the so-called 
XFEL laboratory is part of the design of the electron-positron linear 
collider TESLA. 
In addition to the immediate applications in condensed matter physics, chemistry, 
material science, and structural biology, 
X-ray FELs may be employed also to study some physics issues of fundamental
nature. In this context, one may mention the boiling of the vacuum 
(Schwinger pair creation in an external field), horizon physics 
(Unruh effect), and axion production.
We review these X-ray FEL opportunities of fundamental physics and 
discuss the necessary technological improvements in order to achieve these goals.  
}

\section{Introduction}

There are definite plans for the construction of free electron lasers (FELs) in the
X-ray band, both at the Standord Linear Accelerator Center (SLAC), where the so-called Linac Coherent Light 
Source (LCLS) has been proposed~\cite{Arthur:1998yq}, 
as well as at DESY, where the so-called 
XFEL laboratory is part of the design of the electron-positron ($e^+e^-$) linear 
collider TESLA (TeV-Energy Superconducting Linear Accelerator)~\cite{Brinkmann:1997nb,Materlik:2001qr}.

X-ray free electron lasers will give us new insights into natural and life sciences.
X-rays play a crucial role when the structural and electronic properties of
matter are to be studied on an atomic scale. The spectral characteristics
of the planned X-ray FELs, with their high power, short pulse length, 
narrow bandwidth, spatial coherence, and a tunable wavelength, make 
them ideally suited for applications in atomic and molecular physics, plasma physics, 
condensed matter physics, material science, chemistry, and 
structural biology~\cite{Arthur:1998yq,Brinkmann:1997nb,Materlik:2001qr}.

In addition to these immediate applications, X-ray FELs may be employed also to study 
some physics issues of fundamental nature~\cite{Tajima:2000}. 
In this context, one may mention the boiling of the 
vacuum~\cite{Melissinos:1998qn,Chen:1998,Ringwald:2001ib,Alkofer:2001ik,Popov:2001ak} 
(Schwinger pair creation in an external field), 
horizon physics~\cite{Chen:1999kp,Rosu:1999} (Unruh effect), and 
axion production~\cite{Melissinos:1999aq,Ringwald:inprep}.
In this contribution, I shall concentrate on these particle physics 
opportunities of X-ray FELs. 
I shall also discuss the necessary improvements in X-ray FEL technology in order to achieve 
these goals.

\section{X-Ray Free Electron Lasers}

Before we enter the discussion of particle physics issues, let us briefly review
the principle of X-ray free electron lasers.

\begin{figure}
\begin{center}
\psfig{figure=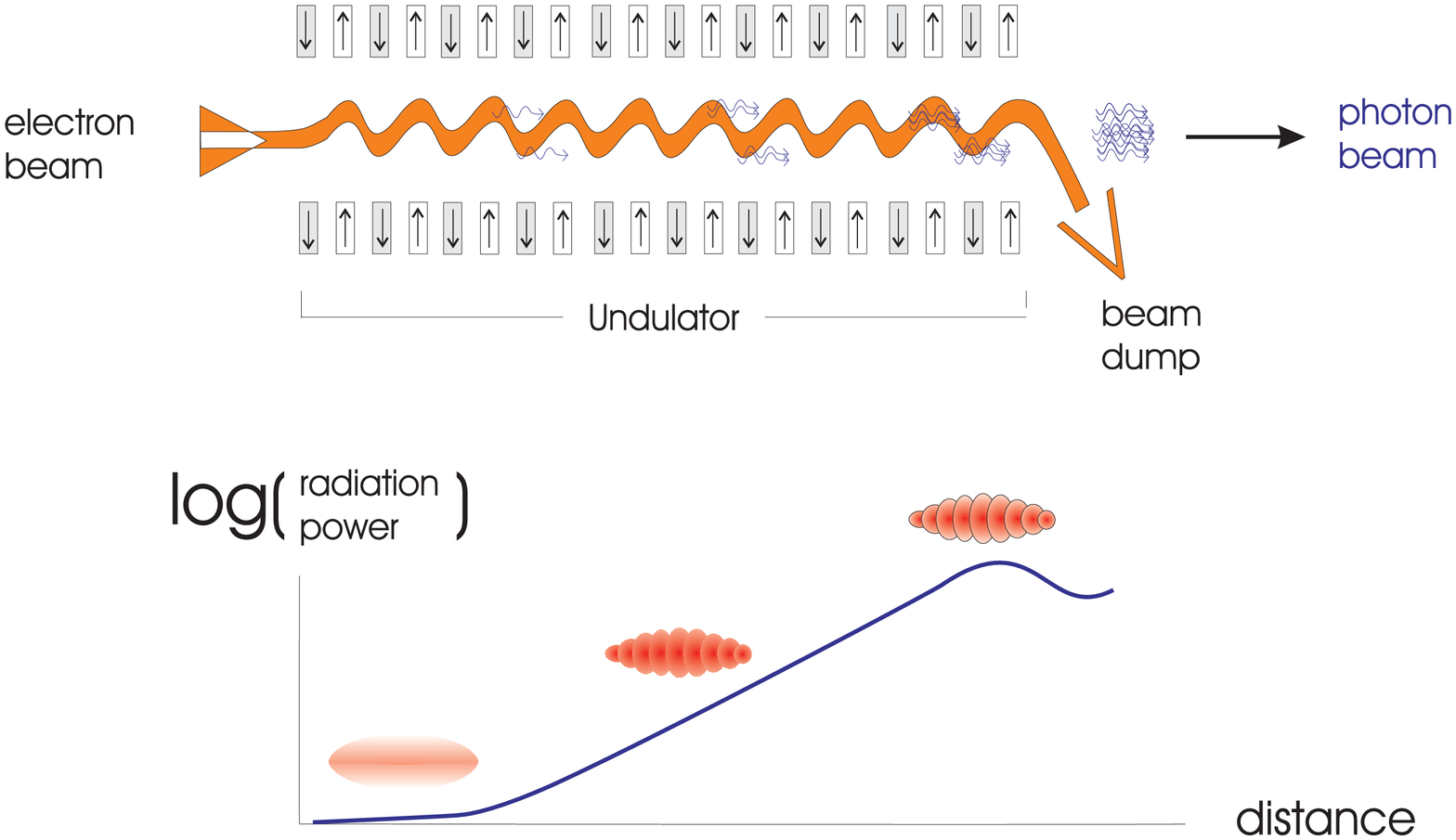,width=11.5cm}
\caption[...]{Principle of a single-pass  X-ray free electron laser in the self amplified spontaneous 
                      emission  mode~\cite{Materlik:2001qr}.
\label{fig:xfel_princ}}
\end{center}
\end{figure}

Conventional lasers yield radiation typically in the optical band. The reason is 
that in these devices the gain comes from stimulated emission from electrons
bound to atoms, either in a crystal, liquid dye, or a gas.
The amplification medium of free electron lasers~\cite{Madey:1971}, 
on the other hand, is {\em free} 
(unbounded) electrons
in bunches accelerated to relativistic velocities with a characteristic longitudinal
charge density modulation (cf. Fig.~\ref{fig:xfel_princ}). The radiation emitted by
an FEL can be tuned over a wide range of wavelengths, which is a very important 
advantage over conventional lasers. 

\begin{figure}
\begin{center}
\psfig{figure=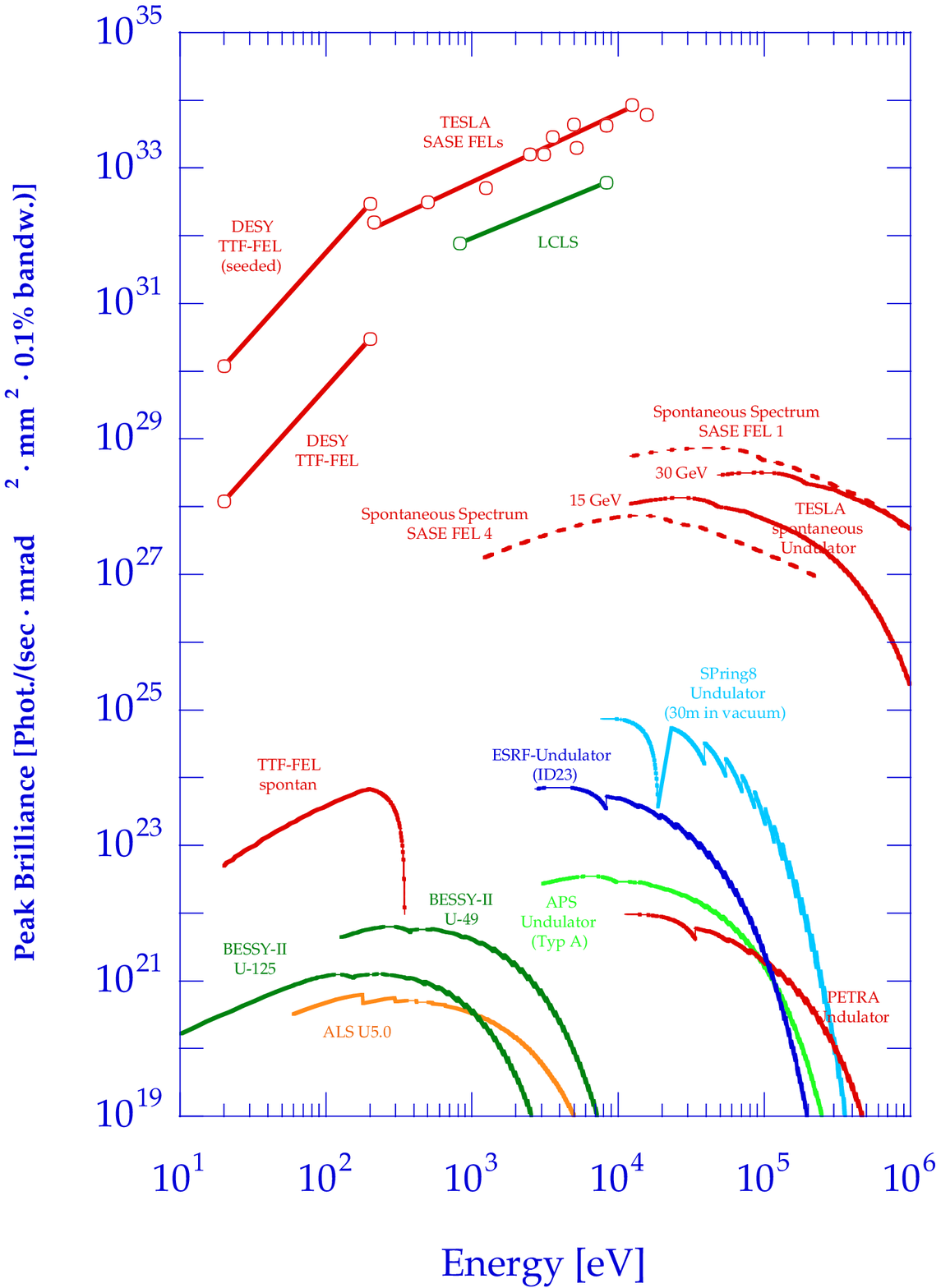,width=8.2cm}
\parbox{7.5cm}{\vspace{-11cm}
\psfig{figure=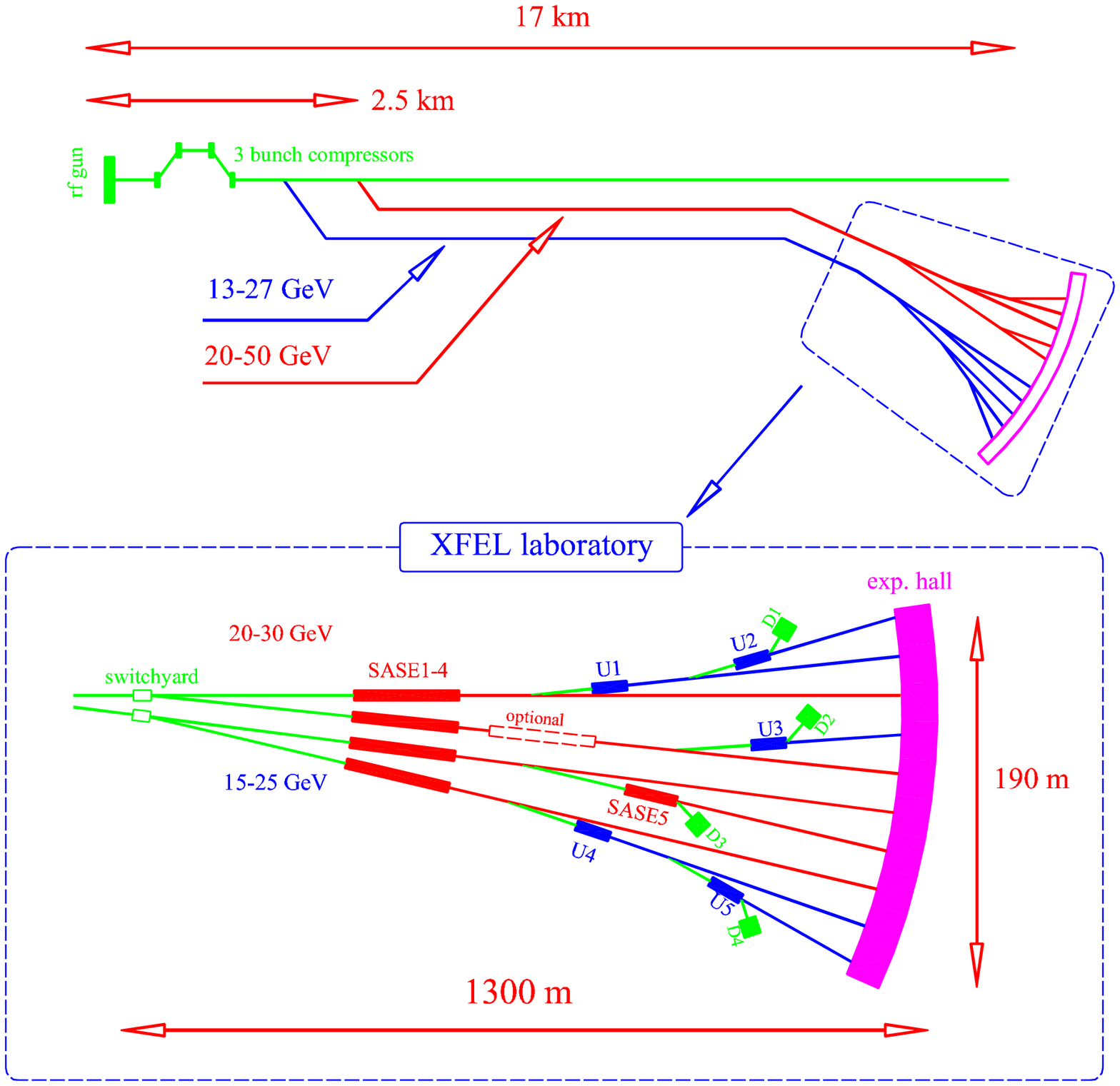,width=7.5cm}}
\caption[...]{{\em Left:} Spectral peak brilliance of X-ray FELs and undulators for 
                      spontaneous radiation at TESLA, together with that of third generation
                     synchrotron radiation sources~\cite{Materlik:2001qr}. For comparison, the spontaneous 
                      spectrum of an X-ray FEL undulator is shown.
                       {\em Right:} Schematic view of the TESLA XFEL electron beam transport (top)
                      and the XFEL laboratory (bottom)~\cite{Materlik:2001qr}. 
\hfill
\label{fig:xfel_spect}}
\end{center}
\end{figure}

The basic principle of a single-pass free electron laser operating in the self amplified
spontaneous emission (SASE) mode~\cite{Kondratenko:1980} is as follows.
It functions by passing an electron beam pulse of energy $E_e$ of small 
cross section and high peak current through a long periodic magnetic structure (undulator)
(cf. Fig.~\ref{fig:xfel_princ}). 
The interaction of the emitted synchrotron radiation, with opening angle
\begin{equation}
     1/\gamma = m_e c^2/E_e 
= 2\cdot 10^{-5}\ \left(  25\ {\rm GeV}/E_e
\right) \,,
\end{equation}
where $m_e$ is the electron mass, 
with the electron beam pulse within the undulator leads to the buildup of a
longitudinal charge density modulation (micro bunching), if a resonance condition,
\begin{equation}  
     \lambda = \frac{\lambda_{\rm U}}{2\gamma^2}
     \left( 1 + \frac{K^2_{\rm U}}{2}\right)
= 0.3\ {\rm nm}\ \left( \frac{\lambda_{\rm U}}{1\,{\rm m}}\right) 
\left( \frac{1/\gamma}{2\cdot 10^{-5}}\right)^2\,
\left( \frac{ 1 + K^2_{\rm U}/2 }{3/2}\right)
\,, 
\end{equation}
is met. Here, $\lambda$ is the wavelength of the emitted radiation, 
$\lambda_{\rm U}$ is the length of the magnetic period of the undulator, and
$K_{\rm U}$ is the undulator parameter,
\begin{equation}
K_{\rm U} 
= \frac{e \lambda_{\rm U} B_{\rm U} }{
                2\pi m_e c}\,, 
\end{equation}
which gives the ratio between the average deflection angle of the electrons in the 
undulator magnetic field $B_{\rm U}$ from the
forward direction and the 
typical opening cone of the synchrotron radiation. The undulator parameter should be
of order one on resonance. 
The electrons in the developing micro bunches eventually radiate
coherently -- the gain in radiation power $P$, 
\begin{equation}
P\propto e^2\,{N_e^2}\,B_{\rm U}^2\,\gamma^2\,,
\end{equation}
over the one from incoherent spontaneous synchrotron radiation ($P\propto N_e$) 
being proportional to the number $N_e\geq 10^9$ of electrons in a bunch 
(cf. Fig.~\ref{fig:xfel_spect} (left)) --  
and the number of emitted photons grows exponentially
until saturation is reached.
The radiation has a high power, short pulse length, narrow bandwidth, is fully polarized, 
transversely coherent, and has a tunable wavelength. 

The concept of using a high energy electron linear accelerator for building an 
X-ray FEL was first proposed for the Stanford Linear Accelerator~\cite{Arthur:1998yq}. 
The feasability
of a single-pass FEL operating in the SASE mode has recently been demonstrated~\cite{Andruszkow:2000it}
down to a wavelength of 80 nm using electron bunches of high charge density and low
emittance from the linear accelerator at the TESLA test facility (TTF) at DESY. 
An X-ray FEL laboratory is planned as an integral part of TESLA~\cite{Brinkmann:1997nb,Materlik:2001qr} 
(cf. Fig.~\ref{fig:xfel_spect} (right)). Some characteristics of the radiation from the  planned 
X-ray FELs at TESLA are listed in Table~\ref{tab:xfel_par}. 

\begin{table}[b]
\caption[...]{Typical photon beam properties of the SASE FELs at TESLA~\cite{Materlik:2001qr}.
\label{tab:xfel_par}}
\vspace{0.4cm}
{\small
\begin{center}
\begin{tabular}{|lc||c|c|c|c|c|}\hline 
 & unit & SASE 1 & SASE 2 & SASE 3 & SASE 4 & SASE 5\\\hline
wavelength  & nm & $0.1\div 0.5$ & $0.085\div 0.133$ & $0.1\div 0.24$ & $0.1\div 1.0$ & $0.4\div 5.8$ \\
peak power & GW & 37 & 19 & 22 & 30 & $110\div 200$\\ 
average  power & W & 210 & 110 & 125 & 170 & $610\div 1100$\\
numb.  photons per pulse & \#  & $1.8\cdot 10^{12}$ &$8.2\cdot 10^{11}$&$1.1\cdot 10^{12}$&$1.5\cdot 10^{12}$&
$2.2\div 58\cdot 10^{13}$\\
bandwidth (FWHM) & \% &0.08&0.07&0.08&0.08&$0.29\div 0.7$\\
pulse duration (FWHM) & fs & 100 & 100 & 100 & 100 & 100
\\\hline
\end{tabular}
\end{center}
}
\end{table}

\section{Applications in Particle Physics}

The spectral characteristics of X-ray free electron lasers suggest immediate
applications in condensed matter physics, chemistry, material science, and structural biology, 
which are reviewed in the conceptual~\cite{Arthur:1998yq,Brinkmann:1997nb} and 
technical~\cite{Materlik:2001qr} design reports of the planned X-ray FEL facilities.   
In this section, we want to emphasize that X-ray FELs may be employed also to study some 
particle physics issues. In this context, one may mention the boiling of the 
vacuum~\cite{Melissinos:1998qn,Chen:1998,Ringwald:2001ib,Alkofer:2001ik,Popov:2001ak} 
(Schwinger pair creation in an external field), 
horizon physics~\cite{Chen:1999kp,Rosu:1999} (Unruh effect), and 
axion production~\cite{Melissinos:1999aq,Ringwald:inprep}.

Whereas the last application mainly requires large average radiation power $\langle P\rangle$, 
the first two applications require very large electric fields
and thus high peak power densities $P/(\pi\sigma^2)$, where $\sigma$ is the laser spot radius. Here, 
one could make use of the possibility to focus the X-ray beam to a spot of   
small radius, hopefully down to the diffraction limit, 
$\sigma\,\gwig\, \lambda\simeq \mathcal{O}(0.1)$~nm. In this way, one may 
obtain very large electric fields and accelerations, 
\begin{eqnarray}
    \label{peak-electric-field}
    { \mathcal E} &= & \sqrt{
    \mu_0\,c\,
    \frac{P}{\pi \sigma^2} }
    \ = \ { 1.1\cdot 10^{17}}\ { \frac{\rm V}{\rm m}}\ 
    \left( \frac{P}{1\ {\rm TW}}\right)^{1/2}\,
    \left( \frac{0.1\ {\rm nm}}{\sigma}\right)\,,
    \\ \label{peak-accel}
    { a} &=& \frac{e\,{\mathcal E}}{m_e} = 
    { 1.9\cdot 10^{28}}\ { \frac{\rm m}{{\rm s}^2}}\ 
    \left( \frac{P}{1\ {\rm TW}}\right)^{1/2}\,
    \left( \frac{0.1\ {\rm nm}}{\sigma}\right) \,,
\end{eqnarray}
much larger than those obtainable with any optical laser of the same power.

\subsection{Boiling the Vacuum}

Spontaneous particle creation from vacuum induced by an external field, first 
put forth to examine the production of 
$e^+e^-$ pairs in a static, spatially uniform electric 
field~\cite{Sauter:1931} and often 
referred to as the Schwinger mechanism, ranks among the 
most 
intriguing nonlinear phenomena in quantum field theory. Its consideration is
theoretically important, since it requires one to go beyond 
perturbation theory, and its experimental observation would 
verify the validity of the theory in the domain of strong fields.  
Moreover, this mechanism has been applied to many problems in 
contemporary physics, ranging from black hole quantum 
evaporation~\cite{Hawking:1974rv}
to particle production in hadronic 
collisions~\cite{Casher:1979wy} 
and in the early universe~\cite{Parker:1969au}, to 
mention only a few. One may consult the 
monographs~\cite{Greiner:1985} for a review of further 
applications, concrete calculations and a detailed bibliography.

It is known since a long time that in the background of a static, spatially 
uniform electric field the vacuum in quantum electrodynamics (QED) is unstable
and, in principle, sparks with spontaneous emission of $e^+e^-$ 
pairs~\cite{Sauter:1931}.
However, a sizeable rate for spontaneous pair production requires 
extraordinary strong electric field strengths $\mathcal E$ of order or
above the critical value
\begin{equation}
{\mathcal E}_c \equiv \frac{m_e\, c^2}{e\, \lambdabar} 
= \frac{m_e^2\, c^3}{e\, \hbar}
\simeq 1.3\times 10^{18}\ {\rm V/m}\,. 
\label{schwinger-crit}
\end{equation}
Otherwise, for $\mathcal E\ll \mathcal E_c$, the work of the 
field on a unit charge $e$ over the Compton wavelength of the electron 
$\lambdabar =\hbar /(m_e c)$ is much smaller than the rest energy 
$2\,m_e c^2$ of the produced $e^+e^-$ pair, the
process can occur only via quantum tunneling, and its rate is  
exponentially suppressed, 
\begin{eqnarray}
\label{schwinger-rate}
 \frac{{\rm d}^4\,n_{e^+e^-}}{{\rm d}^3x\,{\rm d}t} \sim  
\frac{c}{4\,\pi^3 \lambdabar^4}\, 
\exp \left[ -\pi\, \frac{{\mathcal E}_c}{\mathcal E} \right]\,.
\end{eqnarray}

Unfortunately, it seems inconceivable to produce macroscopic 
static fields 
with electric field strengths of the order of the Schwinger critical 
field~(\ref{schwinger-crit}) in the laboratory. In view of this difficulty,    
in the early 1970's the question was raised\footnote{At about the same time, 
the thorough investigation of the question started whether the necessary 
superstrong fields around ${\mathcal E}_c$ can be generated microscopically and
transiently in the Coulomb field of colliding heavy ions with 
$Z_1+Z_2 > Z_c\approx 170$~\cite{Zeldovich:1972}. At the present 
time, clear experimental signals for spontaneous positron creation in heavy 
ion collisions are still missing and could
only be expected from collisions with a prolonged lifetime~\cite{Greiner:1998}.}  
whether intense optical lasers could be employed to 
study the Schwinger mechanism~\cite{Bunkin:1970,Brezin:1970}. Yet, it was 
found that all available and conceivable optical lasers did not have enough 
power density to allow for a sizeable pair creation rate~\cite{Bunkin:1970,%
Brezin:1970,Popov:1971,Troup:1972,%
Popov:1974,Mostepanenko:1974,Popov:1974b,Marinov:1977gq,Katz:1998,Dunne:1998ni,Fried:2001ga}. 

\begin{table}[t] 
\caption[dum]{ 
Laser parameters and derived quantities relevant for 
estimates of the rate of spontaneous $e^+e^-$ pair production~\cite{Ringwald:2001ib}. 
The column labeled ``Optical'' lists parameters which are typical for
a petawatt-class (1 PW = $10^{15}$ W) optical laser~\cite{Perry:1994}, 
focused to the diffraction limit, 
$\sigma = \lambda$. The column labeled ``Design'' displays design 
parameters of the planned XFEL at DESY 
(``SASE-5'' in Ref.~\cite{Materlik:2001qr} and Table~\ref{tab:xfel_par}). Similar values apply for 
LCLS~\cite{Arthur:1998yq}.
The column labeled ``Focus: Available'' shows typical values which can
be achieved with present day methods of X-ray 
focusing~\cite{Graeff:priv,Hastings:2000}: It assumes that the XFEL 
X-ray beam can be focused to a rms spot radius of $\sigma \simeq 21$ nm with 
an energy extraction efficiency of 1 \%~\cite{Graeff:priv}. The column 
labeled ``Focus: Goal'' 
shows parameters which are theoretically possible by increasing the energy 
extraction of LCLS (by the tapered undulator technique) and by a yet 
unspecified method of diffraction-limited focusing of X-rays~\cite{Chen:1998}.
\hfill
\label{parameters}} 
{\scriptsize
\begin{center}
\begin{tabular}{|ll|c||c|c|c|}\hline 
\multicolumn{6}{|c|}{{\bf Laser Parameters}}\\\hline
 & & {\bf Optical}~\cite{Perry:1994}  & \multicolumn{3}{c|}{\bf X-ray FEL}
\\\hline
 & & Focus: & Design~\cite{Materlik:2001qr} & Focus: & 
Focus: \\  
   & & Diffraction limit &  & Available~\cite{Graeff:priv} & 
Goal~\cite{Chen:1998}  
\\\hline 
Wavelength& $\lambda$ & 1 $\mu$m & 0.4 nm & 0.4 nm & 0.15 nm \\
Photon energy & $\hbar\,\omega = \frac{hc}{\lambda}$ & 1.2 eV & 3.1 keV 
& 3.1 keV 
& 8.3 keV\\
Peak power& $P$ & 1 PW & 110 GW  & 1.1 GW &  5 TW                \\
Spot radius (rms)& $\sigma$ & 1 $\mu$m & 26 $\mu$m & 21 nm  & 0.15 nm\\
Coherent spike length (rms) & $\triangle t $ & 500 fs $\div$ 20 ps
& 0.04 fs
& 0.04 fs
& 0.08 ps               \\\hline
\multicolumn{6}{|c|}{{\bf Derived Quantities}}\\\hline
 & & & & & \\[0.5ex]
Peak power density&$S=\frac{P}{\pi \sigma^2 }$& $3\times 10^{26}$ 
$\frac{\rm W}{{\rm m}^2}$
& $5\times 10^{19}$ $\frac{\rm W}{{\rm m}^2}$ & 
$8\times 10^{23}$ $\frac{\rm W}{{\rm m}^2}$ & $7\times 10^{31}$ 
$\frac{\rm W}{{\rm m}^2}$
\\[1ex]
Peak electric field&$\mathcal E =\sqrt{\mu_0\,c\, S}$
& $4\times 10^{14}$ $\frac{\rm V}{\rm m}$ & 
$1\times 10^{11}$ $\frac{\rm V}{\rm m}$ &
$2\times 10^{13}$ $\frac{\rm V}{\rm m}$ & 
$2\times 10^{17}$ $\frac{\rm V}{\rm m}$
\\[1ex]
Peak electric field/critical field
 & $\mathcal E/{\mathcal E}_c$  &$3\times 10^{-4}$ & $1\times 10^{-7}$   
& $1\times 10^{-5}$  & 0.1     
\\
Photon energy/$e$ rest energy  &$\frac{\hbar\omega}{m_e c^2}$ & 
 $2\times 10^{-6}$ & $0.006$ &
$0.006$ & $0.02$ \\
Adiabaticity parameter & $\eta = 
\frac{\hbar \omega}{e\, \mathcal E \lambdabar}$ 
& $9\times 10^{-3}$ & $6\times 10^{4}$ &
$5\times 10^{2}$ & 0.1 \\
\hline
\end{tabular}
\end{center}
}
\end{table}

With the possibility of X-ray lasers at the horizon, this question has been
addressed recently again~\cite{Melissinos:1998qn,Chen:1998,Ringwald:2001ib,Alkofer:2001ik,Popov:2001ak}. 
As a quasi-realistic picture of the 
electromagnetic field of a laser, a pure electric field oscillating with a 
frequency $\omega = 2\pi c/\lambda$ was considered\footnote{Such a field may be created in an
antinode of the standing wave produced by a superposition of two laser beams. 
In Ref.~\cite{Fried:2001ga}, pair creation in the overlap region of two lasers, whose
beams make a fixed angle to each other, was considered.}, under the assumption that 
the field amplitude $\mathcal E$ is much smaller than the Schwinger critical field,
and the photon energy is much smaller than the rest energy of the electron, 
\begin{equation}
{\mathcal E}\ll {\mathcal E}_c\,,\hspace{6ex}
\hbar\omega\ll m_e c^2\,;
\label{conditions}
\end{equation}
conditions which are well satisfied at realistic X-ray lasers 
(c.\,f. Table~\ref{parameters}). 
Under these conditions, it is possible to compute the rate of 
$e^+e^-$ pair production in a semiclassical manner, using 
generalized WKB or imaginary-time 
methods~\cite{Brezin:1970,Popov:1971,Popov:1974,Popov:1974b,Dunne:1998ni,Fried:2001ga}. 
Here, the ratio $\eta$ of the energy of the laser photons over 
the work of the field on a unit charge $e$ over the Compton wavelength of the 
electron,
\begin{equation}
\label{gamma}
\eta = \frac{\hbar\omega}{e {\mathcal E} \lambdabar }
=
\frac{\hbar\,\omega}{m_e c^2}\,\frac{{\mathcal E}_c}{\mathcal E} =
\frac{m_e c\,\omega}{e\,\mathcal E}\,,
\end{equation}
plays the role of an adiabaticity parameter. As long as $\eta\ll 1$, 
i.\,e. in the strong-field, low-frequency limit, the non-perturbative 
Schwinger result~(\ref{schwinger-rate}) for a static uniform field applies.
On the other hand, for large $\eta$, i.\,e. in the low-field, high-frequency
limit, the essentially perturbative result 
\begin{equation}
 \frac{{\rm d}^4\,n_{e^+e^-}}{{\rm d}^3x\,{\rm d}t} \sim
\frac{c}{4\,\pi^3 \lambdabar^4}\, 
\left(
\frac{{\rm e}}{4}
\frac{eE\lambdabar}{\hbar \omega}
\right)^{4m_ec^2/\hbar\omega}
\label{brezin-rate}
\end{equation}
is obtained for the rate of pair production.
It corresponds to the 
$n$-th order perturbation theory, $n$ being the minimum number of quanta
required to create an $e^+e^-$ pair: $n\gwig 
2\, m_e c^2/(\hbar\omega)\gg 1$.

\begin{figure}
\begin{center}
\psfig{figure=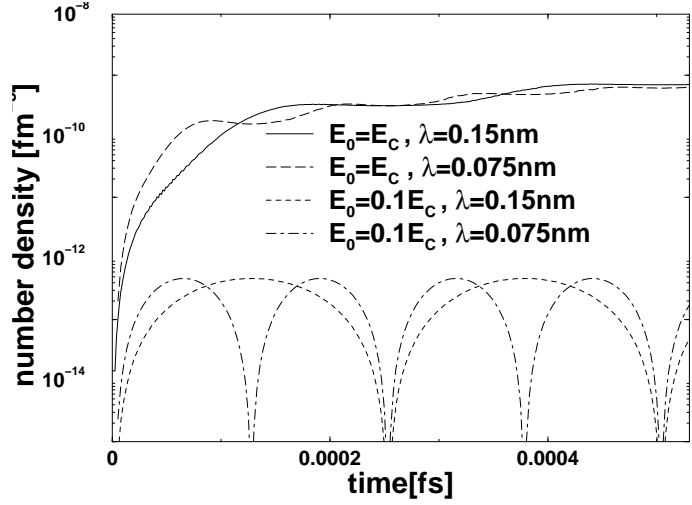,angle=-90,width=9cm,clip=}
\caption[...]{Time evolution of the number density of produced
$e^+e^-$ pairs at the focus of an X-ray laser~\cite{Alkofer:2001ik}. 
\hfill
\label{fig:numb_dens}}
\end{center}
\end{figure}

For an X-ray laser, with $\hbar\omega\sim 1\div 10$ keV, 
the adiabatic, nonperturbative, strong field regime, $\eta\,\lwig\, 1$, 
starts to apply for 
$\mathcal E\,\gwig\, \hbar\omega\, 
{\mathcal E}_c/(m_ec^2)\sim 10^{15\div 16}$ V/m 
(c.\,f. Eq.~(\ref{gamma})). 
An inspection of the tunneling rate~(\ref{schwinger-rate})
leads then to the conclusion that   
one needs an electric field of about $0.1\,{\mathcal E}_c\sim 10^{17}$ V/m 
in order to get an appreciable amount of  spontaneously produced $e^+ e^-$ pairs 
at an X-ray laser~\cite{Ringwald:2001ib}. Under such conditions  
the production rate is time-dependent, with repeated cycles of 
particle production and annihilation in tune with the laser frequency 
(cf. Fig.~\ref{fig:numb_dens}), but 
the peak particle number is independent of the laser frequency: up to 
$10^3$ pairs may be produced in the spot volume~\cite{Alkofer:2001ik}.  

In Table~\ref{parameters} we have summarized the relevant parameters for the 
planned X-ray FELs. We conclude that 
the power densities and electric fields which can be reached with presently
available technique (column labeled ``Focused: Available'' in Table~\ref{parameters})
are far too small for a sizeable effect. On the other hand, if the  
energy extraction can be improved considerably, such that 
the power of the planned X-ray FELs can be increased to the terawatt region,
and if X-ray optics can be improved~\cite{Hastings:2000} to approach the diffraction limit of focusing,
leading to a spot size in the 0.1 nanometer range, then there is ample
room (c.\,f. column labeled ``Focus: Goal'' in Table~\ref{parameters}) for an 
investigation of the Schwinger pair production mechanism at 
X-ray FELs. 

\subsection{Unruh Effect}

Black hole evaporation, the so-called Hawking effect~\cite{Hawking:1974rv}, and the 
similar  
Unruh effect~\cite{Unruh:1976db} are two fundamental effects in present-day
theoretical physics. Both are thermal-like effects involving 
microscopic degrees of freedom of quantum fields which are not 
causally connected (event horizons).  

Experimental detection of Hawking radiation from real, massive ($M_{\rm bh}$) 
general-relativistic black holes seems 
impossible, 
since the corresponding temperature, as seen by an outside stationary observer, is tiny,
\begin{equation}
T_{\rm Hawking} 
= \frac{\hbar\,\kappa}{2\,\pi\,k}
= 6\cdot 10^{-8}\ {\rm K}\ \left( \frac{1\, M_\odot}{M_{\rm bh}}\right)\,,
\end{equation}
where $\kappa$ is the surface gravity of the black hole, 
$k$ is the Boltzmann constant, and $M_\odot$ denotes the solar mass. 
Furthermore, the proposed detection~\cite{Halzen:uw} of Hawking radiation 
from primordial mini black holes, which are relics from the big bang, is rather 
indirect~\footnote{In models of TeV-scale quantum gravity with extra 
dimensions~\cite{Arkani-Hamed:1998rs}, 
mini black holes may be generated and their evaporation studied at 
near-future collider~\cite{Giddings:2001bu} and existing cosmic ray facilities~\cite{Feng:2001ib}.}. 
Even the detection of condensed matter analogues of Hawking radiation, while more accessible than 
that from real black holes, is currently far from laboratory realization 
(see e.g. Ref.~\cite{Rosu:1994ic} and references therein). 

Under these circumstances, it seems worthwhile to look more closely onto the Unruh effect.
It implies that a  particle uniformly accelerated by an acceleration $a$ will find
itself surrounded by a thermal heat bath at a temperature
\begin{equation}
\label{eq:unr_temp}
T_{\rm Unruh}=
\frac{\hbar\,{ a}}{2\,\pi\, c\, k} =
{ 4\cdot 10^{-21}\ {\rm K}}\  \left( \frac{ a}{1\ {\rm m}/{\rm s}^2}\right)\,.
\end{equation}
We see that enormous accelerations are
required to produce a measurable effect. Here, the X-ray lasers come into play: 
very large accelerations might be available at their focus (cf. Eq.~(\ref{peak-accel})). 

\begin{figure}[t]
\begin{center}
\psfig{figure=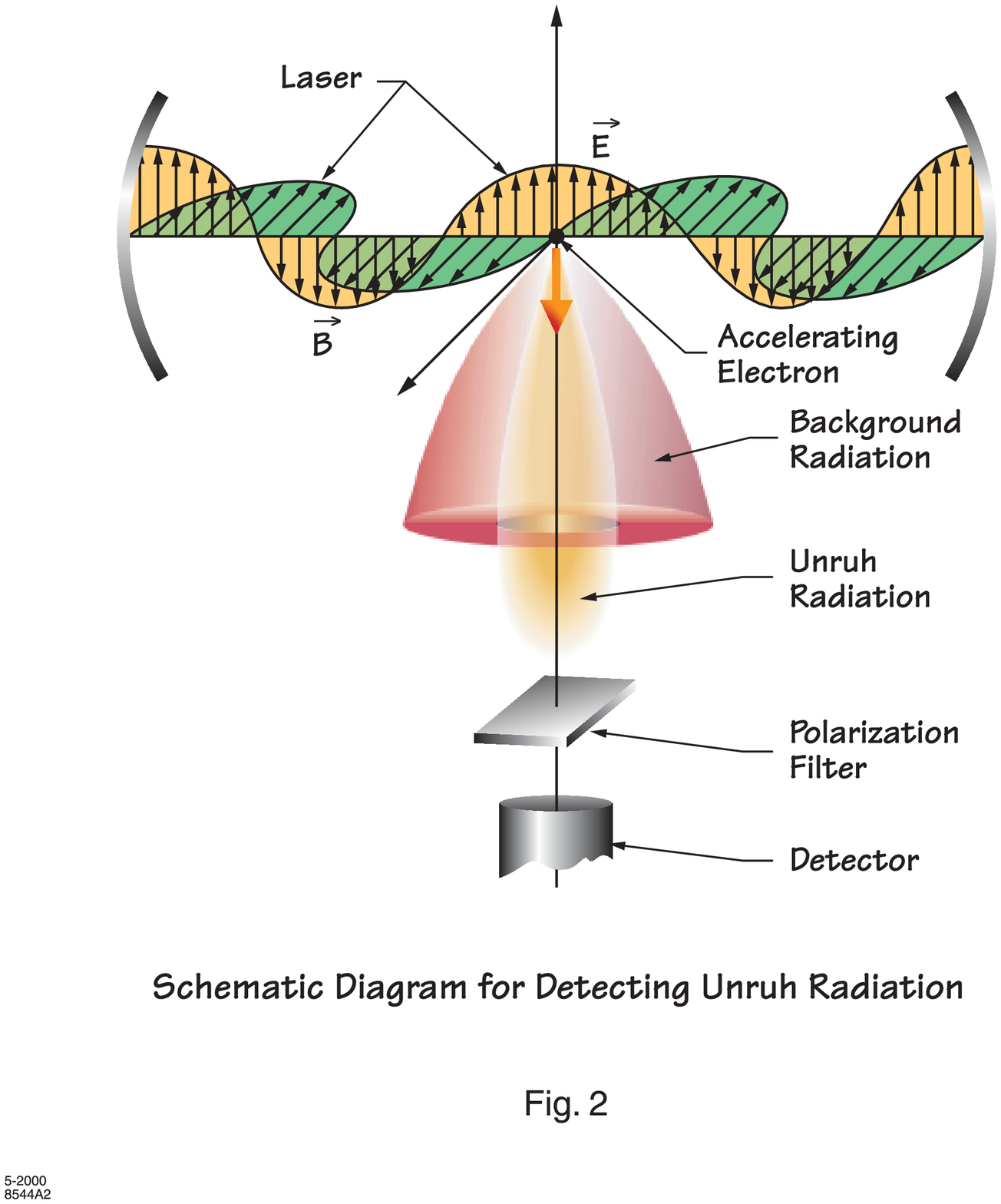,bbllx=70pt,bblly=170,bburx=530pt,bbury=636pt,width=6cm,clip=}
\caption[...]{Schematic diagram of an experiment to detect the Unruh effect at an X-ray 
free electron laser~\cite{Chen:priv}.
\hfill
\label{fig:unr_det}}
\end{center}
\end{figure}

A scheme how the Unruh effect could be detected at the focus of an X-ray 
laser~\cite{Chen:1999kp,Rosu:1999} 
is shown in Fig.~\ref{fig:unr_det}. In the spot of a standing laser wave, where only an
electric field exists, a single electron is accelerated with $a\approx 10^{26}$ m/s$^2$, 
a value which is possible with state-of-the-art means of focusing (cf. Eq.~(\ref{peak-accel})
and Table~\ref{parameters}, column labeled ``Focus: Available'').   
The acceleration of the
electron through the vacuum causes a jitter in the electron's motion, in addition to
the usual zero point fluctuations. This jitter modifies the radiation emitted by
the electron -- over and above the classical Larmor radiation. The additional, acceleration-related
radiation has a characteristic $a$ dependence (a distorted thermal spectrum) and 
angular dependence (cf. Fig.~\ref{fig:unr_det}). 
In particular, there is a blind spot in the angular dependence of the classical
Larmor radiation. Any radiation in this blind spot should be traceable to the distortion
of the zero-point fluctuations.    

Whether ultimately one will call this a verification of the Unruh effect or just basic
quantum field theory (QED) is a matter of taste or linguistics~\cite{Rosu:1999}.
After all, the Unruh temperature itself will not be measured. 
Nevertheless, it seems worthwhile to pursue this type of experiment.

\subsection{Axion Production}

An axion ($A^0$)~\cite{Weinberg:1978ma} is a hypothetical, 
very light, weakly coupled (pseudo-)scalar particle.
It arises from a natural solution to the strong $CP$ problem: why the effective
$\theta$-parameter in the Lagrangian of quantum chromodynamics (QCD), 
\begin{equation}
{\mathcal L}_\theta = 
\theta_{\rm eff}\, \frac{\alpha_s}{8\pi}\, G^{\mu\nu a} 
\tilde G_{\mu\nu a}\,,
\end{equation}
is so small, $\theta_{\rm eff}\lwig 10^{-9}$, as
required by the current limits on the neutron electric dipole moment, even 
though $\theta_{\rm eff}\sim 1$ is perfectly allowed by QCD gauge invariance?
Here, $\alpha_s$ is the strong fine-structure constant, 
and $G^{\mu\nu a}$ ($\tilde G_{\mu\nu a}$) are the (dual) gluon field strength tensors. 
The axion appears as a pseudo Nambu-Goldstone boson of a spontaneously brocen Peccei-Quinn
symmetry~\cite{Peccei:1977hh}, whose scale $f_A$ determines the mass,
\begin{equation}
\label{eq:ax_mass}
{ m_A} = { 0.62\cdot 10^{-3} \  {\rm eV}}\  
         \times
         \left( 
         \frac{10^{10}\ {\rm GeV}}{ f_A} 
        \right)\,.
\end{equation}
The original axion model, with $f_A\sim v = 247$ GeV being of the order of 
the scale of electroweak symmetry breaking, is experimentally excluded (see e.g. Ref.~\cite{Groom:2000in}
and references therein), however so-called invisible axion models~\cite{Kim:1979if,Zhitnitsky:1980tq}, 
where $f_A\gg v$, are still allowed.    

\begin{figure}[t]
\begin{center}
\psfig{figure=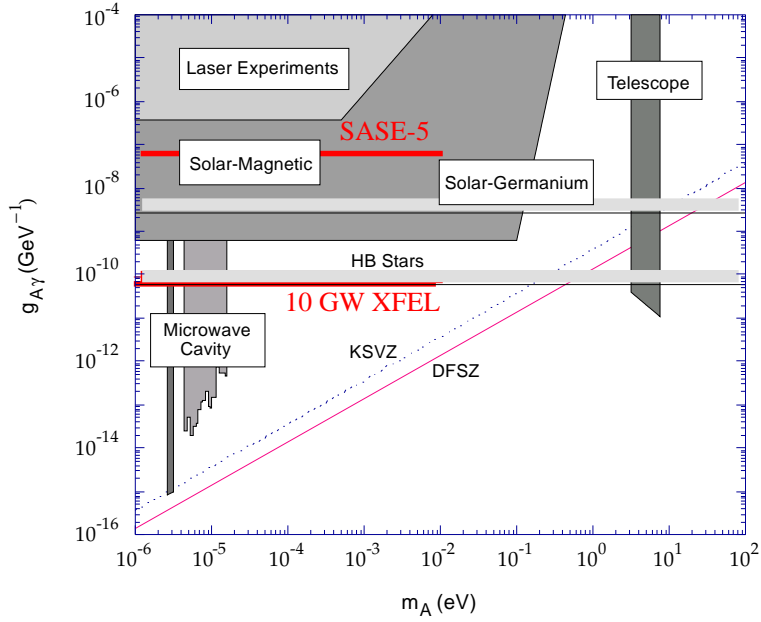,width=10cm}
\caption[...]{Exclusion region in mass $m_A$ vs. axion-photon coupling $g_{A\gamma}$ for various 
current experiments (adapted from Ref.~\cite{Groom:2000in}, where also the corresponding references 
can be found).  
Also shown in this figure, and labeled with 
``SASE-5'', is the projected sensitivity~\cite{Ringwald:inprep} of a photon regeneration 
experiment using the SASE-5 XFEL (cf. Table~\ref{tab:xfel_par}), as well as the one 
of a hypothetical XFEL with average power $\langle P\rangle = 10$ GW (``10 GW XFEL''). 
\hfill
\label{fig:ax_ph}}
\end{center}
\end{figure}

The interactions of axions with Standard Model particles are model dependent. Of particular importance
is the axion-photon coupling $g_{A\gamma}$, 
\begin{equation}
\label{eq:ax_ph}
        {\mathcal L}_{\rm WW} = 
\frac{1}{4}\,{ g_{A\gamma}}\,A^0\ F_{\mu\nu} \tilde{F}^{\mu\nu} 
=
- { g_{A\gamma}}\,A^0\ {\mathbf E}\cdot {\mathbf B}\, ;
        \hspace{5ex}
        { g_{A\gamma}} = \frac{\alpha}{2\pi { f_A}} 
        \left( { \frac{E}{N}} -1.92\right) \,,
\end{equation}
where $F_{\mu\nu}$ ($\tilde{F}^{\mu\nu}$) is the (dual) electromagnetic field strength tensor.
The quantity $E/N$ is the ratio of electromagnetic over color anomalies, a model-dependent
ratio of order one. It is noteworthy however, that 
two quite distinct models, namely the so-called KSVZ~\cite{Kim:1979if} (or hadronic) 
and the so-called DFSZ~\cite{Zhitnitsky:1980tq} (or grand unified) one, lead to quite similar
axion-photon couplings, as shown in Fig.~\ref{fig:ax_ph}, 
which displays the 
axion-photon coupling~(\ref{eq:ax_ph}) , in terms of its mass~(\ref{eq:ax_mass}).
Also shown in Fig.~\ref{fig:ax_ph} are the exclusion regions arising from astrophysical
considerations and various experiments\footnote{For a discussion and references, see Ref.~\cite{Groom:2000in}.}. 
Apart from the laser experiments quoted in Fig.~\ref{fig:ax_ph}, all the others 
rely on axion production in cosmological or astrophysical environments and 
aim only at their detection: The microwave cavity experiments assume that axions are 
the galactic dark matter, the telescope search looks for axions thermally produced in 
galaxy clusters, and the solar-magnetic and solar-Germanium experiments search for 
axions from the sun.    

\begin{figure}[t]
\begin{center}
\psfig{file=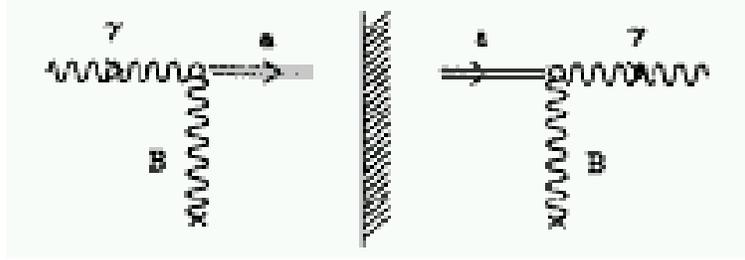,width=10cm,clip=}
%
\caption[...]{Schematic view of axion production through photon conversion 
in a magnetic field (left) and subsequent detection through photon regeneration 
(right)~\cite{Cameron:1993mr}. 
\hfill
\label{fig:light_wall}}
\end{center}
\end{figure}

In connection with the X-ray laser, a photon regeneration experiment seems to be  appropriate
and possible~\cite{Ringwald:inprep}. It may be based on the idea~\cite{Anselm:1985gz}
to send a laser beam along a superconducting dipole magnet (with $\mathbf E\, ||\, \mathbf B$),
in whose magnetic field the photons may convert via 
a Primakoff process (cf. Eq.~(\ref{eq:ax_ph})) into axions (cf. Fig.~\ref{fig:light_wall}). 
If another dipole magnet is set up in line with the first magnet, with a sufficiently thick
wall between them, then photons may be regenerated from the pure axion beam
in the second magnet and detected.  
For light axions, with 
\begin{equation}
\label{eq:ax_coh}
{ m_A} \ll  \sqrt{\frac{4\pi\omega}{\ell}}
=
{ 1.6\cdot 10^{-3}\ {\rm eV}}
       \ 
\,\left( \frac{\hbar{ \omega}}{1\ {\rm eV}} 
       \frac{1\ {\rm m}}{\ell} \right)^{1/2}\,,
\end{equation}
where $\ell$ is the length of the magnetic field, 
the axion beam produced is collinear and coherent with the 
photon beam, and the overall rate for production and subsequent detection is 
\begin{equation}
\label{eq:ax_reg_rate}
{\rm  rate} \propto 
\frac{1}{16} 
       \left( { g_{A\gamma}}\, { B\, \ell} \right)^4
\,          { \frac{\langle P\rangle }{\omega}}
\,,
\end{equation}
where $\langle P\rangle$ is the average laser power. 
A pilot experiment~\cite{Ruoso:1992nx,Cameron:1993mr} with an optical laser, 
$\lambda = 514$ nm, of power $\langle P\rangle = 3$ W, and with a magnet with 
$B = 3.7$ T and $\ell = 4.4$ m, excluded axion-like pseudo-scalars with
mass $m_A<10^{-3}$ eV and axion-photon couplings $g_{A\gamma}>6.7\cdot 10^{-7}$ 
GeV$^{-1}$. The overall envelope of current limits from laser-based experiments
is shown in Fig.~\ref{fig:ax_ph} (``Laser Experiments''). 

Also shown in this figure, and labeled with 
``SASE-5'', is the projected sensitivity~\cite{Ringwald:inprep} 
(cf. Eqs.~(\ref{eq:ax_coh}) and (\ref{eq:ax_reg_rate})) of a possible photon regeneration 
experiment using the SASE-5 XFEL (cf. Table~\ref{tab:xfel_par}), along with 
a state-of-the-art magnetic field of $B = 10$ T over $\ell = 10$ m. 
Clearly, in order to reach a sensitivity comparable to the one arising from 
stellar evolution\footnote{The 
Cern Axion Solar Telescope~\cite{Fanourakis:2001kp} (CAST) has a designed
sensitivity of $g_{A\gamma}\,\gwig\, 5\cdot 10^{-11}$ GeV$^{-1}$ and will compete with the stellar evolution
limit.}, which excludes $g_{A\gamma}\,\gwig\, 6\cdot 10^{-11}$ 
GeV$^{-1}$ and is labeled ``HB Stars'' in Fig.~\ref{fig:ax_ph},  additional efforts are necessary. 
For example, one might envisage an XFEL with $\langle P\rangle =10$ GW, and a magnetic field
of $B= 40$ T over $\ell = 40$ m, to obtain a projected sensitivity of 
$g_{A\gamma}\,\gwig\, 8\cdot 10^{-11}$ GeV$^{-1}$, for $m_A\,\lwig\, 10^{-2}$ eV.

\section{Conclusions}

We have considered several particle physics applications of X-ray free 
electron lasers, such as spontaneous $e^+e^-$ pair creation from vacuum, 
the Unruh effect, and axion production. We have seen that for all these
applications still some improvement in X-ray FEL technology over the 
presently considered design parameters is necessary. 
But these opportunities appear to be very well worth the effort. 
In addition, I should point out that the subject of fundamental physics
at an X-ray FEL is still in its infancy and in a development phase.
No doubt, there will be unpredecedented opportunities to use these intense
X-rays in order to explore even more issues of fundamental physics 
that have eluded man's probing so far.

\section*{Acknowledgments}

I would like to thank H. Mais, A.~Melissinos, 
G.~Raffelt, and J.~Ro\ss bach 
for fruitful discussions and a careful reading of the manuscript. 
Furthermore, I would like to congratulate Bill Marciano and Sebastian White 
for organizing such a remarkable workshop.

\end{document}